# Mobile Computing in Physics Analysis - An Indicator for eScience

Arshad Ali[*], Ashiq Anjum[**], Tahir Azim[*], Julian Bunn[***], Ahsan Ikram[*],
Richard McClatchey[**], Harvey Newman[***], Conrad Steenberg[***], Michael Thomas[***], Ian Willers[****]

[*]National University of Sciences and Technology, Rawalpindi, Pakistan
{arshad.ali, ahsan, tahir}@niit.edu.pk
[**]CCS Research Centre, University of the West of England, Bristol, UK,
{ashiq.anjum, richard.mcclatchey}@cern.ch
[***]California Institute of Technology, Pasadena, CA 91125, USA,
{conrad,newman, thomas}@hep.caltech.edu, Julian.Bunn@caltech.edu
[****]CERN, CH-1211, Geneva 23, Switzerland Ian.Willers@cern.ch

## ABSTRACT

This paper presents the design and implementation of a Grid-enabled physics analysis environment for handheld and other resource-limited computing devices as one example of the use of mobile devices in eScience. Handheld devices offer great potential because they provide ubiquitous access to data and round-the-clock connectivity over wireless links. Our solution aims to provide users of handheld devices the capability to launch heavy computational tasks on computational and data Grids, monitor the jobs' status during execution, and retrieve results after job completion. Users carry their jobs on their handheld devices in the form of executables (and associated libraries). Users can transparently view the status of their jobs and get back their outputs without having to know where they are being executed. In this way, our system is able to act as a high-throughput computing environment where devices ranging from powerful desktop machines to small handhelds can employ the power of the Grid. The results shown in this paper are readily applicable to the wider eScience community.

## 1   INTRODUCTION

Handheld computing and wireless networks hold a great deal of promise in the fields of ubiquitous data access and in particular for the eScience community. However, the analysis and processing of information to produce useful, filtered results is a much larger challenge, since it requires not just network connectivity, but considerable processing power as well. Most mobile and handheld devices, and even many laptop and desktop machines are extremely limited in the processing power they offer. Many tasks are themselves so CPU-intensive and time-consuming that a single machine seems quite insufficient for them. Grid computing is poised to become the technology that will address this limitation. We present here a brief description of a Grid-enabled analysis environment for handheld devices that we are developing specially for the high-energy physics (HEP) community. In this paper, we describe the architecture, design, and implementation of this Grid-enabled system for the handheld and mobile devices. Such an architecture is readily adaptable to other data-intensive eScience analyses.

The CMS (Compact Muon Solenoid) experiment [1] at CERN, due to commence in 2007, will use Grid-based data stores for the gigabytes of data it will generate each minute. In its raw form, the data cannot be used to generate significant results, because of its sheer quantity and complexity. The only way of analyzing this data is by using analysis applications to render it in the form of 2D and 3D diagrams, examinable on handheld devices, which scientists can use much more effectively in understanding the physics taking place in CMS. At the same time, efficient processing of data is required in order to make the analysis process as fast and as interactive as possible.

Our aim in this work is to develop a set of physics analysis applications for handhelds and to optimize them for maximum performance on the handheld devices. At the same time, we have been developing a Java-based framework, called JClarens, for hosting Web and Grid services. The JClarens framework (described in [2]) allows users a single point of access to Grid services, such as data storage and replication services, monitoring services and job submission services. Using JClarens, users can search for data on the Grid and can launch analysis jobs on the datasets from handheld and mobile devices. We not only use the handheld devices as a powerful data centric visualization tool but we also monitor the progress of the Grid jobs from these devices thereby helping the users to access and monitor the information and devices "on-the-fly".

This research aims at making the power of Grid computing available to resource-limited devices such as laptops and PDA's, especially the Pocket PC and Palm. At the same time, we have been working on porting popular physics analysis applications to PDAs. Due to the processing power and resources generally required by physics applications, none of the applications has yet been ported to the low resource (usually 32MB RAM) and slow processing (typically 200MHz to 400MHz) handheld devices. Moreover, the slow, unreliable and sometimes intermittent nature of wireless connections has been a concern. Combining mobile computing and Grid computing should allow mobile and PDA users to submit jobs on the Grid and access its processing which is of great potential interest in the field of distributed scientific applications.

This paper is organised as follows: in the next section we propose a Grid and mobile computing architecture which is suited for distributed physics analysis. This is followed by discussion of both the server-side and client-side design aspects of this architecture in sections 3 and 4. Thereafter we consider related work and outline the future direction of our work before drawing conclusions.

## 2 ARCHITECTURE OF THE ANALYSIS ENVIRONMENT

The analysis environment, shown in Figure 1, consists of two entirely decoupled components. On one side are the resource-limited handheld devices and the applications specifically designed for these devices; on the other side is the JClarens Grid service host, which provides the facilities offered by the Grid to the handheld devices. The clients comprise simple, portable applications for handheld and desktop devices, which communicate with JClarens, using SOAP/XML-RPC. They can be programmed in C++, Java or any other language supporting XML-RPC. Once logged on to JClarens, the clients can access the services offered by it. A detailed overview of the services offered by Clarens, and the way its clients communicate with it is described in [3]. As noted earlier the server basically acts as a Grid service host. It hosts any services and methods that have to be deployed on the Grid communicating with clients using the lightweight XML-RPC protocol. This allows clients to be made extremely simple, and abstracts away all the complexity of the hosted Grid services from the client.

To provide Grid functionality on the handheld device, a separate service for job submission has been implemented in JClarens. This service (hosted on the resource broker) receives job submission requests from clients, and then attempts to determine the most suitable farm available for job submission, using monitoring information received from all the connected farms (or standalone computers). Once this has been located, a job ID is assigned to that particular job, the job submission request is forwarded to that particular farm, and a record of where the request was forwarded is stored in a database. Any subsequent requests for checking the job status, for killing the job or for retrieving its outputs are forwarded to the JClarens server on the farm concerned. Once a job request is forwarded to a particular farm, the JClarens server on that farm creates a temporary staging directory for Condor on the server machine. The executable, required libraries, a dynamically generated submit file and input files are copied into the staging directories, and "condor_submit" is called to submit the job to the Condor pool on the farm. The clients can subsequently check the status of their jobs (which can be running on any one of the available farms) at any time, without having to know where the jobs are actually being executed. Once the jobs are complete, the user can retrieve the results of his jobs at any time, using the relevant XML-RPC calls.

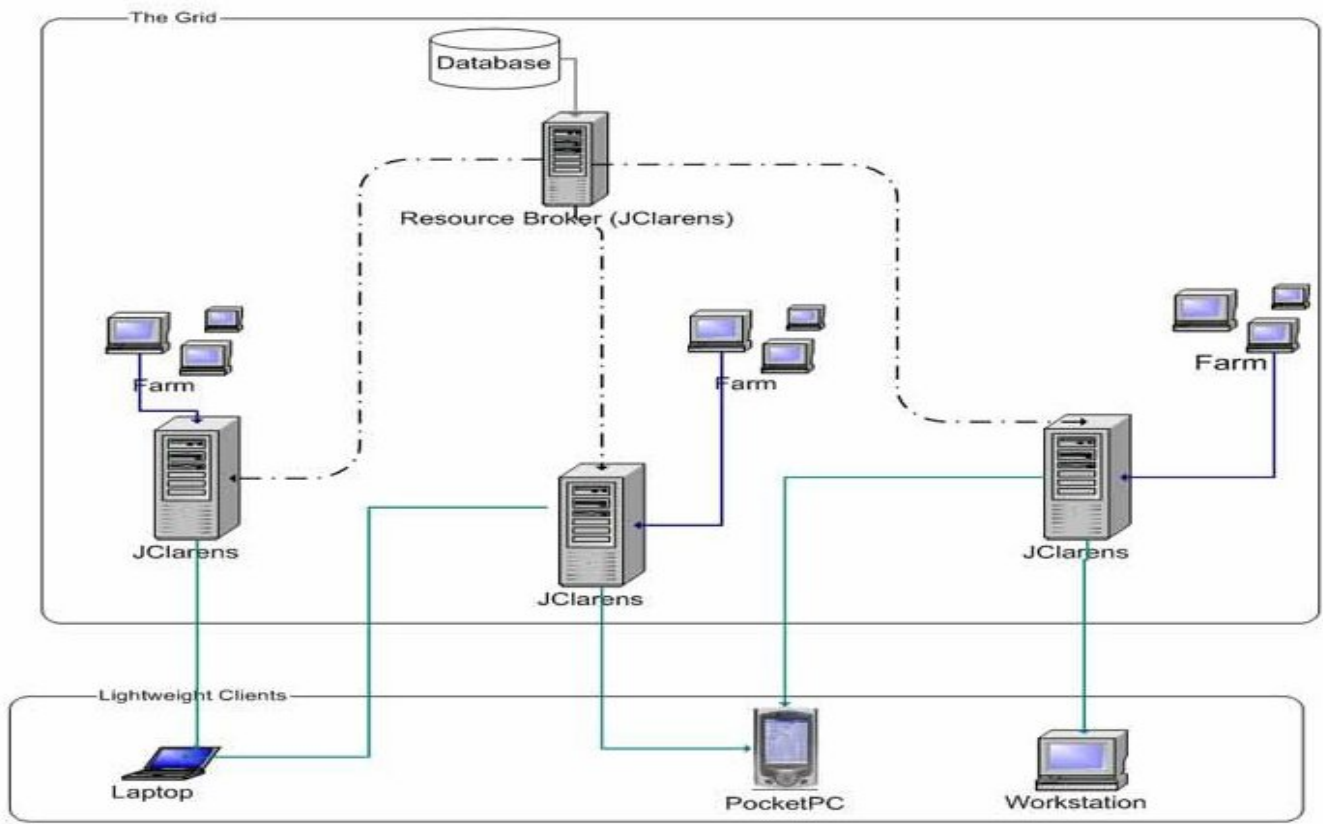

Figure 1: An architectural overview of the analysis environment

# 3 HANDHELD AND MOBILE APPLICATIONS FOR THE GRID

The capability and utility of the system has been validated using handheld applications that have been developed for the Personal Java runtime environment on iPAQ Pocket PCs. The clients we have developed can be run on any mobile device that supports Java. Once installed users can use these applications irrespective of any underlying networks; wireless, wired or mobile. These clients were mainly developed for physicists and scientists to access and analyze data anytime on mobile devices (but of course are more generally applicable in other eScience applications). Later, considering the enormous size of data which is often required by scientists for analysis, we proposed to distribute storage and computation resources on the Grid. In this way we maintained the ubiquitous access and analysis of data; we also saved low resource mobile devices from enormous storage and computation. However, these clients can be easily customized for any solution that requires heavy computation and storage from mobile devices. A description of the clients developed and the features offered by them for interactive analysis follows.

## 3.1 Java Analysis Studio(JAS) & JASOnPDA

The Java Analysis Studio (JAS) has been developed at the Stanford Linear Accelerator (SLAC) [4]. JAS is a physics analysis tool used for analyzing data obtained from linear accelerators in the form of 1D-2D histograms as shown in figure 2. Apart from 1D-2D analysis, JAS offers numerous other facilities, which include comparison of displayed histograms with predefined mathematical functions (Quadratic, Cubic, Gaussian, polynomial, Lorentzian etc), the fitting of these functions over the displayed histograms for statistical analysis (peak value, average value etc), and executing individual analyses code on selected datasets.

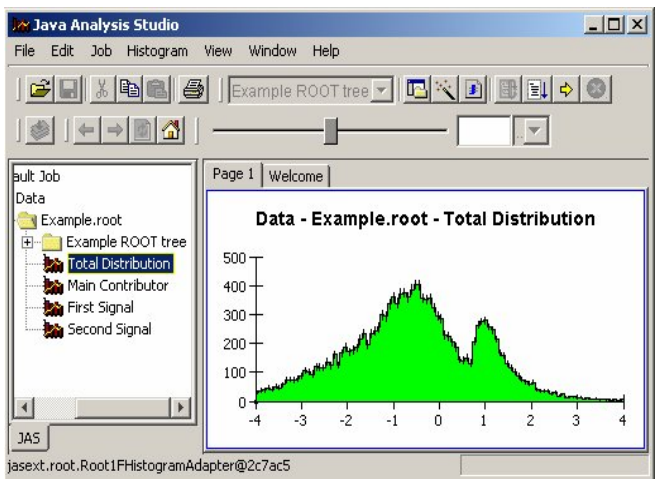

Figure 2: Java Analysis Studio (JAS) (above) running on a Desktop Machine.

JASOnPDA [5] was our first application for the PocketPC. JASOnPDA is the scaled down version of Java Analysis Studio, especially designed for constrained handheld devices. JASOnPDA provides the essential analysis utilities of Java Analysis Studio on PocketPC devices, and was developed using J2SE v1.1. JASOnPDA allows mobile users to log on to the JClarens server using a certificate-based authentication procedure. Once successfully authenticated, the user is allowed to access files stored at the server. The remote browsing facility allows users to browse the directories served by JClarens and to look for desired ROOT files. The selected ROOT [6] file is analyzed and a tree structure displays the hierarchy of objects in the ROOT file. The user can move along the tree, selecting any object from the tree structure, and the selected object will be displayed in the form of 1D-2D Histograms in the display panel as shown in figure 3.

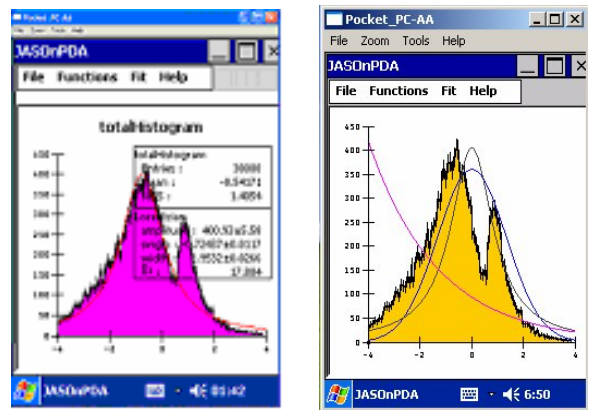

Figure 3: JASOnPDA running on a PDA showing features of histogram plotting, fitting, and statistics calculation.

In order to submit jobs to the Grid using JClarens, JASOnPDA allows users to select the file that will be used as the job's executable. It also allows users to type in or to select a file that will be used as the submit file. Once this has been carried out, the user can select the input files that will be used as the input for the jobs, and can finally submit the job. A notification is received when the job is successfully submitted, after which JASOnPDA periodically polls JClarens for the jobs' status. As soon as the job is complete, a menu is displayed showing the files on which the job execution was successful. The user can then select a particular file to get the resulting output of the job execution on that particular file. In this way, users on handheld devices can submit jobs on compute farms, and at the same time, get back outputs of the job execution. Statistical fitting features from JAS have also been ported. The user can also view statistics information related to the histogram displayed on the screen. Keeping in view the small screen size of the PocketPC, different viewing options are also provided.

To overcome the issue of intermittent, unreliable connections during the transfer of large datasets and files, the downloading process for large files is carried out by dividing the size of the file into small chunks, and downloading those small file chunks in steps, rather than using a single connection to transfer the entire file. This allows us to checkpoint the file transfer process and to ensure that if there is a disconnection at any stage, the entire dataset is not lost, and downloading can be resumed from the latest checkpoint.

Recently, more features for extensibility have been added. Interfaces have been exposed that allow users to write their own analysis classes for different file formats. This provides users the flexibility to select the procedure by which their file will be analyzed and to have its contents displayed. To specify the class to be used for analysis, the user only has to give the name of the class in a simple properties file. In this way, the user can easily plug in the classes that (s)he wants to use for handling new file formats and user-specific custom file formats.

### 3.2 WWW Interactive Remote Event Display (WIRED) and WiredOnPDA

The WWW Interactive Remote Event Display (WIRED) [7] was a joint venture between Stanford Linear Accelerator (SLAC) and the European Organization for Nuclear Research (CERN). WIRED is one of the first physics event displays written in Java for use on the World-Wide-Web. It provides a framework for writing event displays as shown in figure 4. It is in active use by the BaBar and GLAST experiments and the LCD detector study at SLAC.

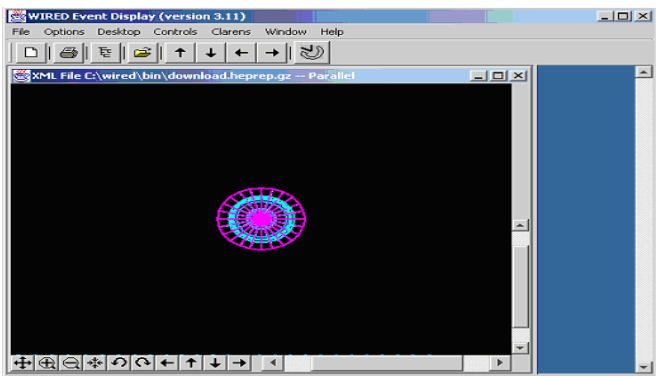

Figure 4: WWW Interactive Remote Event Display running on a Desktop machine.

WiredOnPDA is another of our analysis applications developed for PocketPC devices. As the name suggests, WiredOnPDA provides analysis features of the WWW Interactive Remote Event Display (WIRED) on a PocketPC and is presented in figures 5 and 6. WiredOnPDA accesses data using JClarens in the same way as JASOnPDA.

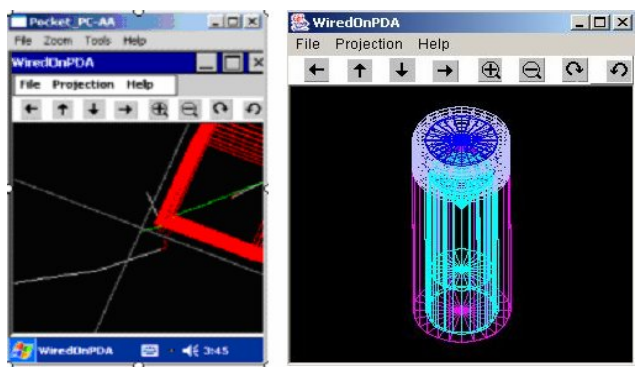

Figure 5: Two views of WiredOnPDA displaying events from a HepRep2 file (left) and detector geometry (right).

As mentioned earlier, the user has to pass a security check by providing a valid certificate and key. Once authenticated, the user can access the remote server and can select any HepRep2 event file placed on the server.

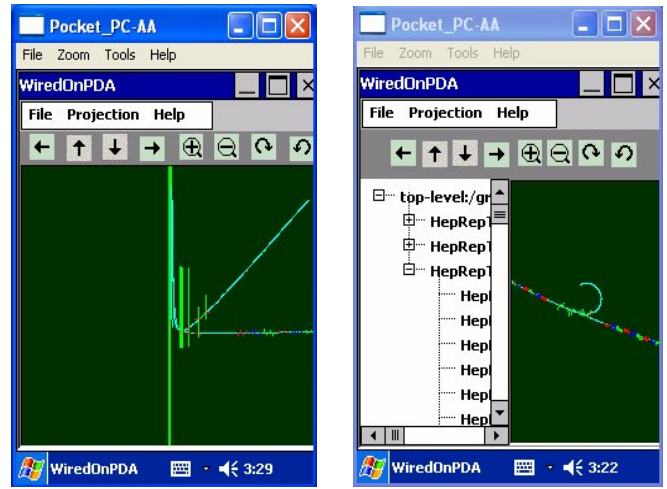

Figure 6: Two views of WiredOnPDA displaying event data and the structure of a HepRep2 file in separate panes.

Once the user selects a file from the JClarens server, the file is downloaded into the RAM and a SAX XML parser [8] parses the information stored in the file. "Drawables" are then extracted from the parsed data and are displayed in a hierarchical tree structure in a WiredOnPDA tree panel. User can then select any "drawable" from the tree and it will be displayed in the display console. Again, keeping in mind the small screen size of the PocketPC, various display options are provided in order to utilize the maximum screen space for event display. As shown in the screenshot figures, the application is provided with a tool bar that allows the user to scale, rotate or zoom the displayed event for improved analysis.

WiredOnPDA also had some issues on its initial release, most of which were regarding performance. The main reason for non-optimal performance was the poor parsing speed of the SAX parsers in PersonalJava. Performance analysis has shown that the reason for this slow parsing is due to the differences in implementation of the PersonalJava virtual machine compared to the J2SE virtual machines. To address this issue various parsers were tried and tested including Xerces [9], Crimson [10], KXML [11] and Piccolo [12]. Piccolo has so far proved to be the fastest performer, with the best possible results amongst the parsers.

### 3.3 Intermittent Connections and Data Transfer

Since wireless networks can be subject to intermittent network connection and loss of data during data transfer we have devised a mechanism to cater for interrupted connections which works as follows. Applications keep a log of the files transferred and the number of bytes of the file that have been transferred. Files were downloaded in N kilobyte chunks into a temporarily-created file, and every n kilobytes, the entry for the file was updated. Once a file is totally transferred, the entry for the file was removed from the log, and the temporary file was deleted. If the connection was disconnected (such as when the timeout expired or the PDA itself detected a broken connection) the user simply has to wait until (s)he gets next connected and then tries to download that file again. However, since the extent of the file downloaded was saved in the temporary file and the number of correctly transferred bytes was stored in the log, the download can be restarted from that number of bytes.

## 4 SERVER SIDE DESIGN

The server side relies mainly on the JClarens framework and the services provided by it. A detailed description of the clients is given in section 3. Here we describe the overall design of the server-side Grid service host, how jobs are actually submitted to the Condor job scheduling system, and how they are managed. A modified version of Apache's XML-RPC [13] classes is used to provide XML-RPC parsing and processing capabilities. The XML-RPC server, encapsulated in a Java servlet, is responsible for processing incoming requests, for extracting information about what services and methods to invoke, and for writing out responses in XML. In addition, we have also modified it to be able to write back responses directly in a binary form, to reduce overheads during binary file transfers.

The major components on the server side are detailed in the following sub-sections.

### 4.1 Authentication and Authorization

Authentication and authorization of valid users and virtual organizations (VOs) is one of the most important requirements of any Grid-enabled system. JClarens carries out authentication using a GSI-based [14] security protocol. The server and clients use X.509 certificates and RSA keys for authentication. A detailed description of Clarens authentication protocol is given in [2]. Authorization is carried out using access control lists (ACLs) maintained in a database. The database system available by default is MySQL [15], although any other DBMS can be easily plugged-in by providing a suitable JDBC driver, and by modifying the configuration settings in a properties file.

The database contains the distinguished names (DNs) or substrings of DNs that are allowed or denied access to various services. A set of methods is also available for manipulating these ACLs by system administrators. Besides users' DNs, the names of the VOs can also be used to configure access control for large groups of users or VOs. JClarens also provides data browsing, searching and downloading capabilities. Using the "file" service, users can browse files, search for files using "wildcards", and search within files. They can also download files and find out the validity of downloaded files using md5 hash values.

### 4.2 Monitoring service

A prototype monitoring service has been developed which reports monitoring information to a central JClarens server, hereafter known as the resource broker. The monitoring service gathers monitoring information using a companion End-host Monitoring Agent (EMA), a monitoring product is also currently being developed. EMA gathers information such as the CPU clock rate, CPU usage, Memory Usage and Last 1, 5 and 15-minute average load values. These parameters are used to calculate a "load coefficient", which describes the load on a particular system in quantitative terms. The coefficient is calculated using an empirical formula:

Coeff. = $((1-CPU\_Usage/100)*-1*Clock\_Rate) + \Sigma a_i M_i$

Where CPU_Usage is the percentage of CPU power being used, $M_i$ is the value of the i'th monitoring parameter and $a_i$ is an experimentally determined weighting factor for the i'th monitoring parameter

Currently the monitoring parameters being used are the percentage of memory being used, the Disk I/O in Mbps, the average CPU load during the past one minute (Load1), and the number of currently executing processes.

Since CPU Usage indicates how much the CPU is being used, (1-CPU_Usage/100) in the first term shows the proportion of the CPU that is free. In addition, because -1 is used as a factor in the term, the more the CPU is free, the lower is the value of the first term. The second term in the formula indicates the load being added to a node because of other factors, like Disk I/O, memory I/O and other factors. In this way, the lower the value of the coefficient for a node, the lower is the load on that node.

Every few seconds (10 by default), this coefficient value is sent to the resource broker as an XML-RPC call, where it is stored in a database, along with the URL of that particular server. This monitoring information is used at the time of job submission to decide where to forward jobs for execution. In order to provide the resource broker more up-to-date (and hence more accurate) monitoring information about a server, the interval after which the data is published to it can be decreased. However, this can result in the generation of more network traffic. To compensate, the network traffic can be reduced at the expense of updated monitoring information by increasing the interval after which the data is published. Therefore, it is up to the administrator of each site to determine a threshold interval that can increase the accuracy of the information published and keep network traffic down to a manageable level.

### 4.3 Job submission service

The job submission service is designed to submit jobs to the least loaded computer or a farm on the network (LAN or WAN). The job submission service receives requests for submission of a job from handheld clients with the following parameters:

1) The name of the job
2) The binary code of the executable
3) The submit file
4) The name of the submit file
5) The names of the input files

The handheld clients themselves do not provide the input files at present. Instead the clients request the JClarens server for files available in its file publishing area and select the ones on which they wish to perform analysis.

On receiving a job submission request (condorSubmit), the receiving JClarens server gets the minimum load coefficient reported by the other peer servers from the resource broker. It then forwards the job to the least loaded peer. Note that the peer is basically a JClarens server running on another computer or on the head node of a compute farm.

When a peer JClarens server receives a forwarded job submission request from another peer, it creates a temporary staging directory for Condor on the local hard disk (see figure 7). It then generates a unique job ID for the request and creates a subdirectory in that folder with the same name as the job ID. A number of subdirectories are then created in the job's particular subdirectory equal to the number of input files and the executable and the submit file are copied into each of the subdirectories, naming them the same as the file names in the request. If the forwarding peer is the local host itself, the server copies the file from the local file system to the appropriate subdirectory directly. Otherwise, it downloads the file from the forwarding JClarens server using the "file.read" call.

Once the staging directories are prepared, JClarens submits the jobs to Condor by executing "condor_submit" calls on the submit file. Condor submits jobs in logical groups called "clusters". A mapping between the jobs themselves and the clusters to which the jobs have been submitted is also stored in a database.

Condor [16] then takes over the allocation of the individual jobs to different machines in the Condor pool and ensures that the jobs are completed and the results returned to the respective directories. While Condor executes the job on the pool, the client can repeatedly poll JClarens to check the status of its jobs. Because only the job ID is required to check the status of a job, a very lightweight XML-RPC request has to be sent to the server to check the status of a job.

In case of successful submission, JClarens returns a unique ID generated for the job, and this is sent to the initial submitting client through the master JClarens. A table mapping the job ID with the peer where the job has been actually submitted is also updated. After that, the client uses this job ID to inspect the status of the job, or to retrieve its status at any later stage.

0. EMA periodically sends monitoring information to the JClarens monitoring interface;
1. The User authenticates, asks for a list of files and then browses the files;
2. The User gets back the list of files, selects files, and downloads the files if required;
3. The User submits jobs;
4. The monitoring information is read by JClarens to select the best place to which the job can be submitted;
5. The Job is forwarded to the best available site;
6. If necessary Input Files are downloaded to the remote execution site;
7. The User checks the status of jobs repeatedly, gets back outputs when desired etc.

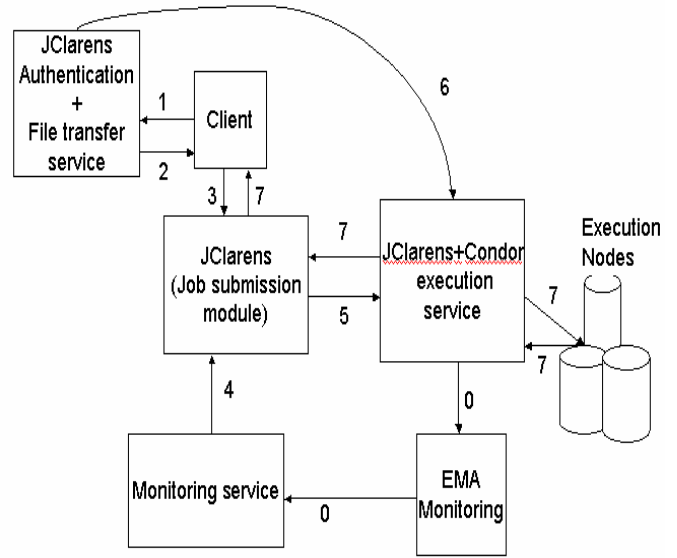

Figure 7: An overview of the steps involved in submitting jobs and checking their status:

JClarens retrieves the status of a job with a particular ID by finding out the peer where the job with that ID has been submitted. The execution of "condor_q" on that peer gives the status of all the jobs running on the machine. The status of the jobs for that particular job ID is extracted by filtering out the clusters mapped to that job ID from the output of "condor_q".

Once the job is complete, the handheld client can retrieve the outputs of all its jobs through JClarens. Clients can look at all the output files produced during execution, and can download the desired output files. In this way, a complex and time-consuming set of jobs can be rapidly executed on an underutilized Condor pool and its results displayed and visualized on the submitting handheld client.

To provide fault tolerance and high availability, the system has been designed to be entirely self-healing. Whenever a new JClarens server initiates and reports to the resource broker, the resource broker informs all the available servers of this new server. In this way if at any stage the central JClarens resource broker goes down, one of the other servers detects this faulty condition automatically, and can establish itself as the central resource broker for all the other available servers. In this way, it is ensured that the system is able to heal itself and remain available for the clients even if the central resource broker becomes inaccessible.

Using this architecture, even users on slow-processing PDAs connected to wireless networks can accomplish complex high end jobs. They can carry out remote processing of lengthy operations on powerful machines in the Condor pool. This allows even handheld devices to exploit the power of Grid computing.

## 5 RELATED WORK

Mobile devices have hitherto not been used for launching and visualizing data intensive applications and jobs. For desktop machines, on the other hand, a variety of Grid enabled physics analysis applications is available. Some of these include JAS [4], ROOT [6] and WIRED [7]. As detailed earlier, JAS is used primarily for 1D and 2D graphical display of histogram data obtained from particle accelerators. Along with graphical display, it also offers various mathematical functions to fit along the displayed data.

WIRED was developed at CERN in collaboration with SLAC. Wired uses XML based files for graphical rendering of events and sub component geometry information from various particle experiments. Also developed at CERN, ROOT analyzes special format ROOT files in which data is arranged in a highly efficient, hierarchical structure.

Research is already being carried out to integrate the two groundbreaking fields of Grid computing and mobile computing 17]. Even Java-enabled mobile phones have been targeted for possible integration with the Grid environment, in order to provide more computationally intensive features on mobile phones [18]. Oracle and Vodafone announced a joint initiative to offer mobile Grid computing [25] in which they would offer enterprise customers integrated mobility solutions based on Oracle 10g and Vodafone Network Services.

Our Pocket PC based analysis applications are basically built around the Java Analysis Studio (JAS) and WWW Interactive Remote Event Display (WIRED) software, and have been optimized for the Personal Java [19] environment on the Pocket PC 2002. On the server side, JClarens is used to provide a framework on which Grid services are hosted to provide Grid authentication, job submission, job tracking, data access, and file browsing services. Another direction that we have researched is in balancing mobile client load by using the JADE-LEAP [20] multi-agent system. In this architecture mobile clients send agents along with job credentials to a central server which then locates a suitable site and dispatches the agent along with the job to that site. After execution the mobile client communicates with another agent residing in the client to transfer results. This approach has yielded useful input to the current research and the work continues in parallel with this effort. The reader is refereed to [21] for further information on multi-mobile agents studies.

San Parak et al. [23] describe the disconnected operation service in mobile Grid computing. They discuss the intermittent connections but provide little information on the issues of the data intensive applications for the mobile devices. Paul and Naian [24] describe an efficient checkpointing technique for mobile Grid computing systems and try to overcome the issues posed by the intermittent connections. EC funded Akogrimo project [28] envisions that Grid services, pervasively available, should eventually meeting the needs of fixed, nomadic and mobile citizens in the 'everywhere at every time in any context' paradigm. But this project is in its initial phases and will take some time to see the results. Chu and Humphrey [25] describe Mobile OGSI.NET which is created to promote resource sharing and collaboration that improves the user experience. Mobile OGSI.NET extends an implementation of Grid computing, OGSI.NET, to mobile devices. This actually deploys the customized Grid toolkit on the mobile device but does not discuss the data handling and presentation issues on the mobile devices. Zhi Wang et al [27] describe the Grid for wireless and adhoc network. They actually test the applications on warless networks but do not take into consideration the Grid issues related to the mobile devices.

## 6 CONCLUSIONS AND FUTURE WORK

The analysis environment presented in this work has been developed to contribute to eScience in general and the physics community in particular all around the world, to help them in their quest for ubiquitous access to data, over wired and wireless links. But it has yet to reach the high performance standards that are so easily achieved in the case of desktop PC-based applications. The slow performance of handheld devices is a major barrier that has to be overcome to achieve true interactivity. However, the use of a distributed job execution and data analysis environment has enabled us to speed up our analysis applications to a great extent. Even if a set of jobs is submitted on a standalone analysis server, users can get back the results of analysis on almost fifteen event files (each event stored in a 1MB file) in less than a minute, whereas a standalone PocketPC could not process more than three or four files per minute.

The distributed architecture of this model makes it unique as it enables mobile clients to analyze data which are too large for them to store or process. It shows that if merged, Grid computing and mobile computing could make use of the features provided by each technology to engender a new era of high resource consuming mobile applications. This can prove to be the first step towards enabling mobile devices to provide functionality and resources similar to desktop PCs.

Our PocketPC based applications were demonstrated in ITU Telecom World 2003 as a part of the "GRID-Enabled Physics Analysis" demonstration. Also the JASOnPDA system was presented at the first Grid Analysis Environment (GAE) workshop at Caltech in June 2005, and attracted a great deal of attention, due to the fact that it was the first physics analysis application ported to the PDA.

WiredOnPDA is expected to gain at least the same amount of attention as its predecessor. Our work on this analysis environment, however, is far from over. Already our current work proves that resource-constrained devices such as the PocketPC can be interfaced with the Grid, and can play a vital role in the realization of the idea of a Grid-Enabled Analysis Environment (GAE). This has clear implications to and application in other areas of Grid-based science analysis. Mobile and ubiquitous computing has yet to reach a mature level, where it can compete with desktop PCs in the kind of applications it can offer. The achievement of the above mentioned goals should prove to be a giant leap towards the attainment of this level of maturity.

There are still some ideas that are yet to be implemented in this project. The most important of these is an even further decentralized architecture, where there is no master server at all. Instead all the servers are true peers of each other and can get monitoring information from a decentralized source. We are currently working on developing a monitoring application that shows network monitoring information between the client and server and the client and other available servers or peer to peer (P2P) nodes. This application will provide data for some of the new features such as bandwidth utilization and data transfer rates. Also, it will serve as a utility application by which users and network administrators can monitor network performance on their mobile handhelds.

A self-organizing neural network (SONN [22]) can be introduced to learn usage and load patterns and thus schedule jobs more efficiently. Another feature that is planned to be added is the ability to keep a catalogue of all files available in all the servers on the network, thereby allowing users to discover any dataset they require without prior knowledge of the server holding that dataset. One major advantage would be the ability to get job outputs back as soon as they are produced rather than having to retrieve them when the entire job is complete. This would be a major advance towards more interactivity. Further work on the handheld client side includes the extension of the analysis environment to run on PocketPC Phone Edition devices and to be able to use fast cellular networks such as GPRS, for data communication with the remote data servers, over the Internet.

**Acknowledgements**

The authors wish to record their appreciation of the support offered by their host institutes and that of the CMS collaboration at CERN, Geneva.